**Title:**

# Machine Learning-Based Reward-Driven Tuning of Scanning Probe Microscopy: Towards Fully Automated Microscopy

**Authors:**

Yu Liu[1*], Roger Proksch[1,2], Jason Bemis[2], Utkarsh Pratiush[1], Astita Dubey[1,5], Mahshid Ahmadi[1], Reece Emery[1], Philip D. Rack[1], Yu-Chen Liu[4], Jan-Chi Yang[4], Sergei V. Kalinin[1,3*]

[1] Institute for Advanced Materials and Manufacturing, Department of Materials Science and Engineering, University of Tennessee, Knoxville, Tennessee, 37996 USA
[2] Oxford Instruments Asylum Research, Santa Barbara, California 93117, USA
[3] Physical Sciences Division, Pacific Northwest National Laboratory, Richland, Washington, 99354, USA
[4] Department of Physics, National Cheng Kung University, Tainan 70101, Taiwan
[5] Institute for Materials Science and Center for Nanointegration Duisburg-Essen (CENIDE), University of Duisburg-Essen, Universitätsstr. 15, 45141 Essen, Germany

* Corresponding author: yliu206@utk.edu, sergei2@utk.edu

**Abstract**

Since the dawn of scanning probe microscopy (SPM), tapping or intermittent contact mode has been one of the most widely used imaging modes. Manual optimization of tapping mode not only takes a lot of instrument and operator time, but also often leads to frequent probe and sample damage, poor image quality and reproducibility issues for new types of samples or inexperienced users. Despite wide use, optimization of tapping mode imaging is an extremely hard problem, ill-suited to either classical control methods or machine learning. Here we introduce a reward-driven workflow to automate the optimization of SPM in the tapping mode. The reward function is defined based on multiple channels with physical and empirical knowledge of good scans encoded, representing a sample-agnostic measure of image quality and imitating the decision-making logic employed by human operators. This automated workflow gives optimal scanning parameters for different probes and samples and gives high-quality SPM images consistently in the attractive mode. This study broadens the application and accessibility of SPM and opens the door for fully automated SPM.

**Introduction**

Scanning probe microscopy (SPM) has revolutionized our understanding of the nanoworld, providing unprecedented insights into the structure and properties of materials at the nanoscale. This powerful technique allows for structural imaging in diverse environments, including ambient conditions, liquids, and vacuum, making it versatile for various applications [1-3]. Over the years, SPM has evolved significantly, building upon the initial contact and non-contact modes [4, 5] to yield a broad array of advanced imaging modes. These advancements have been driven by continuous improvements in hardware and electronics, culminating in the ability to achieve atomic resolution imaging even in ambient conditions [6].

Topographic imaging in SPM not only reveals the surface morphology but also provides additional information through various signal channels. For instance, in the basic tapping mode with periodic drive, phase signals can be obtained and offer insights into material properties and interactions [7, 8]. Moreover, SPM's versatility is further enhanced by the introduction of multi-frequency modes, as demonstrated in [9, 10]. These modes enable the simultaneous acquisition of multiple data channels, enriching our understanding of the sample's properties.

The range of functional imaging modes available in SPM is broad and encompasses various physical properties, such as magnetic [5, 11, 12], mechanical [4, 5], biological [3, 13-15], ferroelectric [16], and electrical characteristics [5, 17]. Each of these modes allows researchers to probe different aspects of the sample, providing a comprehensive view of its functional behavior. However, the effectiveness and reliability of these functional imaging modes are critically dependent on the quality of topographic imaging. Operators must simultaneously achieve multiple objectives, including optimizing image stability, enhancing resolution, and minimizing both reversible and irreversible changes in the probe and sample states. Achieving these goals is a complex and time-consuming process that requires significant expertise. The success of SPM imaging heavily depends on the operator's skill and experience, often leading to irreproducible results when different operators or probes are involved. For example, when imaging hard samples like crystals and ceramics [3, 18-20], a set of inappropriate scanning parameters can lead to fast probe degradation and poor image quality. On soft samples such as soft matters and liquid samples [2, 14, 15, 21-23], non-optimal scanning parameters lead to irreversible sample damage and poor image quality. This challenge is particularly pronounced when dealing with soft systems, nanoparticles, molecules, high-resolution imaging, and imaging in liquids. Transitioning from robust calibration on hard samples to these more delicate and complex systems requires meticulous optimization.

The foundational imaging modality in SPM is the tapping mode [24, 25], also known as intermittent contact (AC) or amplitude-modulated (AM) mode. This imaging mode accounts for most imaging studies in ambient or liquid, and also serves as foundation for more complex techniques such as KPFM, MFM, EFM, and force-distance curve mapping. In tapping mode, the cantilever is driven at or near its lowest resonance frequency. The amplitude and phase of its oscillations are detected using a lock-in amplifier. As the tip approaches the surface, the tip-surface interactions dampen the cantilever's amplitude, which can then be used as a feedback signal for topographic imaging. A key advantage of tapping mode is its ability to significantly reduce tip and surface damage, making it suitable for imaging much softer samples compared to contact mode.

Optimizing the scanning parameters in tapping mode is traditionally difficult and strongly sample-dependent, requiring extensive human input and learning. Interestingly, this task is ill-suited to machine learning methods, as all surfaces are different and hence the classical supervised ML paradigm is poorly applicable [26]. This problem is equally complex for classical control methods, as it is not clear how the cantilever dynamics translate into image quality. Each sample's unique topography, mechanical properties, and environmental interactions contribute to a complex, high-dimensional object space. Combined with the multiple control parameters available in modern instruments, the imaging process defies straightforward modeling or control. Traditional approaches struggle with this variability, often resulting in suboptimal imaging conditions that can lead to poor resolution, increased noise, or even damage to the sample and the probe [27, 28].

At the same time, we note that human operators learn to optimize the scanning conditions, and this knowledge is transferable between different materials and instruments. Skilled operators can intuitively adjust parameters based on their knowledge and experience, achieving high-quality imaging through iterative experimentation and fine-tuning. Hence, we seek to develop an approach that allows transferring the human operator decision-making principles to machine learning workflows. We pose that human expertise can be represented as a reward function that distills the combination of human heuristics and physical knowledge. By incorporating human expertise via a reward-driven machine learning framework, we automate the optimization process, ensuring consistent, high-quality imaging across diverse samples and conditions, while reducing the dependency on operator expertise and minimizing the risk of errors.

In this work, we demonstrate that reward-driven, rather than data-driven, machine learning methods can effectively automate the optimization of SPM imaging for various samples, probes, and microscopes in a traceable manner with minimal requirement of computation power (e.g. local CPU) that can be fulfilled on a local computer without use of graphic processing unit (GPU), which makes it widely accessible. The key innovation in our approach is the introduction of a reward function that transforms the ill-defined instrument problem into a classical optimization task. Essentially, the reward function emulates human expertise, guiding the machine learning algorithm to optimize imaging parameters and consistently achieve the best possible image quality. By leveraging machine learning, we aim to standardize and streamline the optimization process, reducing dependency on operator expertise and improving the reproducibility of SPM imaging. We also pose that the community-wide development of the reward functions offers an approach for democratization and introduction of FAIR principles in automated experiments.

**Reward function-based optimization**

The key aspect of any microscopy experiment is the optimization of image conditions, which typically represents a considerable fraction of the instrument time and effort of the human operator. During this process, the operator tunes the parameters while simultaneously assessing the quality of the data streaming from the instrument. This tuning of imaging conditions in any microscopy mode can be generally represented as an optimization problem within the parameter space of the instrument controls, which are typically well known and provided by the manufacturer. However, defining image quality presents a very complex problem. For atomically resolved imaging in electron microscopy [29] or scanning tunneling microscopy [30], the shape of atomic columns or the presence and intensity of lattice peaks in Fourier transforms offer natural optimization targets [31]. With the optimization target and parameter space defined, this becomes a classical optimization problem.

In non-atomically resolved imaging, unambiguous measures of image resolution or quality are absent. Furthermore, these cannot be defined using classical machine learning approaches since each topographic image is different and hence can be out of distribution with respect to previous examples. While supervised ML workflows can be built for mature fields such as semiconductors or other manufacturing, this paradigm is limited for open-ended research problems. Additionally, descriptive statistics-based measures tend not to be robust with respect to noise and scan instabilities. Previous attempts at automatically tuning SPM have focused on adjusting the PI (proportional-integral) feedback parameters by defining reward functions based on the alignment of the trace and retrace of the height channel. However, these approaches usually result in scanning parameters that lead to rapid probe and sample degradation.

Despite these challenges, human operators learn to optimize scanning conditions through experience acquired on different samples. They intuitively adjust parameters based on real-time feedback from the instrument and their accumulated knowledge of how different settings affect image quality. This ability to adapt and fine-tune parameters through iterative experimentation enables skilled operators to achieve high-quality imaging even under varying and challenging conditions.

Therefore, the key challenge in automating the optimization of the tapping mode (TM) is to find a reward function that works universally across different probes, materials, topologies, and operational environments. Here we design such a reward function based on trace and retrace scan lines that incorporates both physics-based knowledge and empirical insights. Compared to reward functions defined based on entire images, those based on scan lines are approximately 100 times faster, depending on the number of pixels in the scan. This significant speed advantage makes the optimization process practical for real-time operation. By focusing on the critical features of scan lines, such as the alignment and consistency between trace and retrace lines, we aim to develop a robust and efficient optimization framework that can consistently yield high-quality imaging across a wide range of conditions.

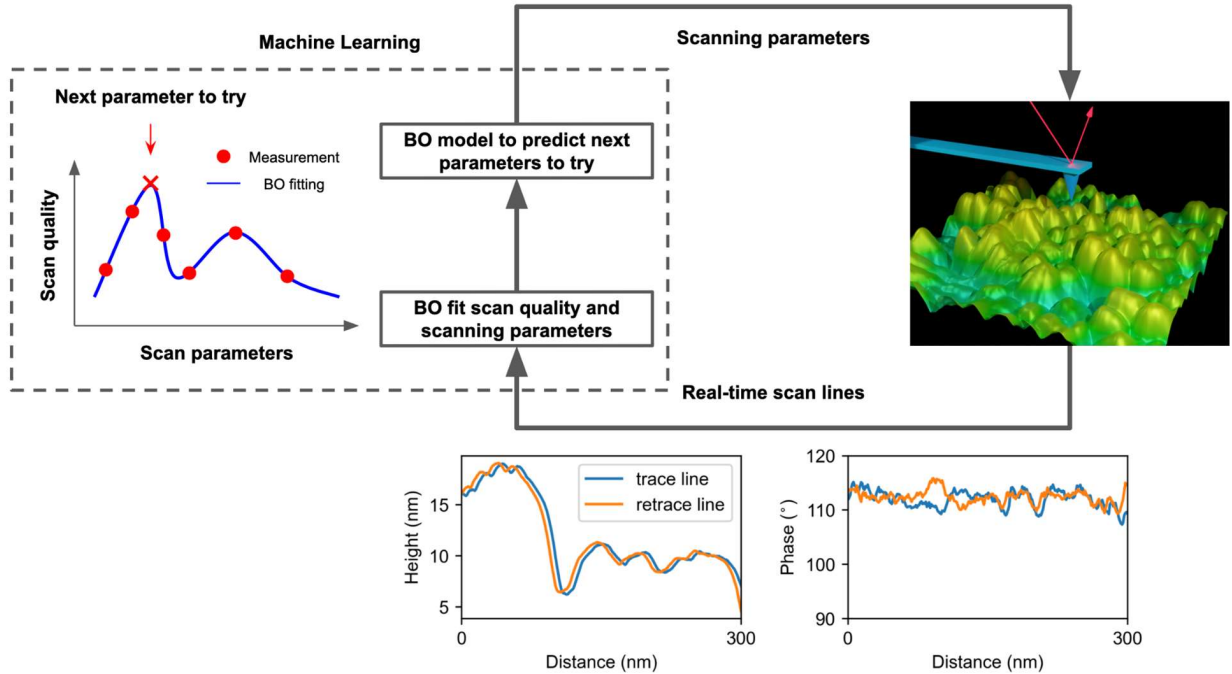

**Figure 1. Workflow of optimizing an autonomous microscope.**-In the autonomous microscope presented here, we define a reward function to quantify the scan quality. During the Bayesian optimization (BO) process, a Gaussian process (GP) model of reward function is fitted for the explored parameters. This GP model is used to predict the distribution and uncertainty of the reward function in the whole parameter space, from which the next set of parameters will be determined. This iteration process is repeated until the maximum number of optimization steps is reached or the image quality is satisfactory. Thus the machine learning algorithm replaces human operation.

**Imaging regimes for TM AFM**

Depending on probe types, samples, and setup parameters, SPM can operate in either attractive or repulsive mode [8]. In attractive mode, the microscope tip operates predominantly under long-range attractive forces, such as van der Waals (vdW) interactions, without physical contact. This allows high-resolution imaging while minimizing deformation and damage of probes and samples. Conditions that favor attractive mode include small amplitudes, larger and less sharp tips, and slower but more sensitive and higher quality factors. Probes often transition from repulsive to mixed to attractive mode as they wear and the balance of vdW and adhesive forces changes. This mode excels in achieving high resolution, preserving the integrity of delicate materials like biomolecules [32], and is ideal for imaging soft or loosely bound materials. However, it requires precise control of the tip-sample distance, necessitating meticulous calibration and stability. In the operation of SPM, attractive mode is usually characterized by phases above the free-air phase, which is the measured phase when the probe is very far away from the sample surface.

In contrast, large amplitudes, sharp tips and lower quality factors favor a net-repulsive interaction - where strong, short range repulsive forces dominate. In repulsive mode, the tip intermittently contacts the sample surface, resulting in predominantly short-range repulsive forces, including mechanical contact and repulsive vdW forces [8]. This mode provides enhanced sensitivity for measuring mechanical properties, such as stiffness and viscoelasticity, and offers better contrast, making it easier to distinguish different materials or features. However, physical contact can lead to sample and probe deformation or damage, reducing image resolution and potentially causing irreversible changes to the probe and sample. While repulsive mode is advantageous for mechanical property measurements, it comes with the trade-off of potential sample alteration and reduced resolution. SPM is operated in the repulsive mode when the measured phases are below the free-air phase.

Because of the tip-shape dependence discussed above, if a probe is initially imaging in repulsive mode, it will often transition from repulsive to mixed to attractive mode as they wear and the balance of vdW and adhesive forces changes. The transition between attractive and repulsive modes can result in high transitory forces [33] and is indicated by phases jumping across the free-air phase.

In this work, we focus on optimizing tip wear while still obtaining high spatial resolution. This requires a stable and robust approach to parameter tuning that can adapt to the varying conditions of different samples and imaging modes. Our goal is to develop a method that ensures consistent high-quality imaging across diverse applications, minimizing the trade-offs between resolution, sensitivity, and sample integrity. By leveraging a reward-driven machine learning framework, we aim to emulate the adaptive expertise of human operators, providing a scalable solution to the complex optimization problem in SPM imaging. This approach promises to enhance the efficiency, reliability and autonomy of SPM, making it more accessible and effective for a wide range of scientific and industrial applications.

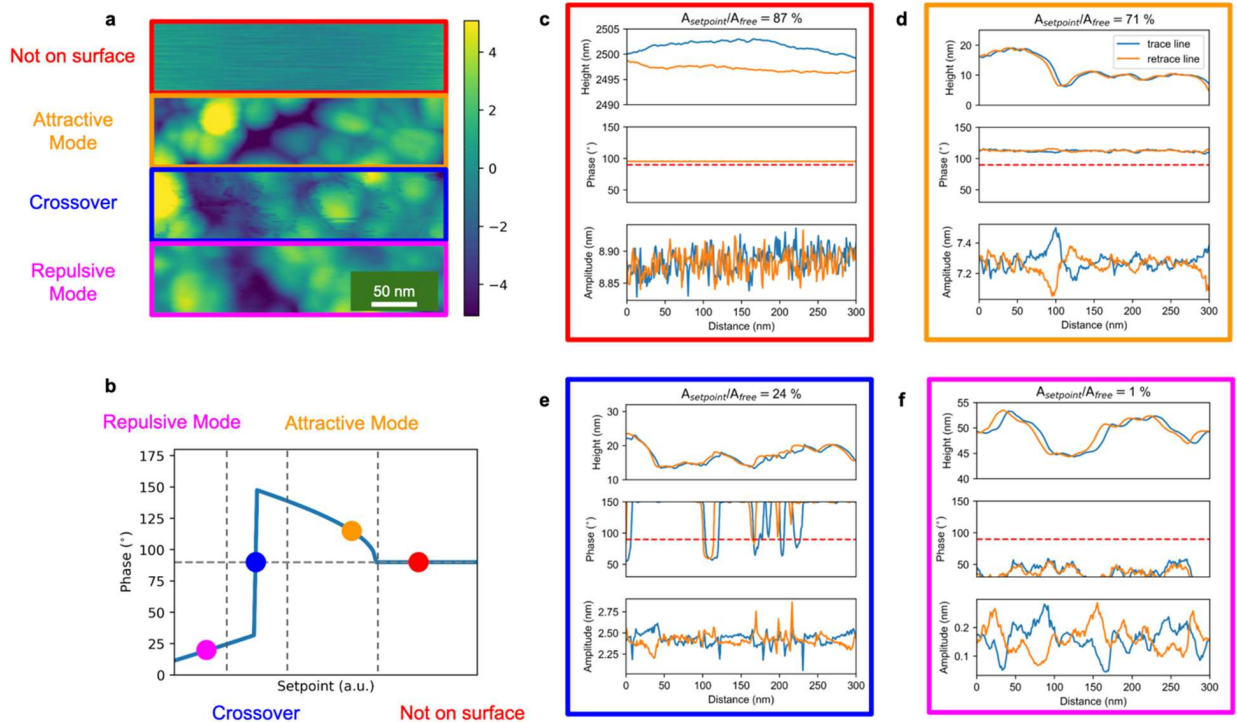

**Figure 2. Define reward function to quantify the image quality. a,** Topography map of $TiO_2$ nanoparticles shows the effect of scanning parameters on the image quality. In the scan, the free-air amplitude is kept at 10.26 nm and the setpoint amplitude is changed in the middle of the scan. **b,** Schematic plot of the dependence of phase on the setpoint at a fixed free-air amplitude. **c,** Scan trace and retrace lines for a large setpoint. When the setpoint is large, the probe is far away from the sample surface and thus dominated by the long-range probe sample interactions. Consequently, the probe follows the profile of this long-range interaction instead of following the sample surface profile closely. **d,** Trace and retrace lines in the attractive mode. In this mode, the system is dominated by net-attractive interactions between probe and sample, which gives high spatial resolution, phase values above 90-degree, and small dissipation that minimizes probe and sample damage. In this study, we are optimizing this imaging mode. **e,** Trace and retrace lines in the attractive-repulsive crossover mode. As the setpoint is lowered, the system alternates between attractive and repulsive modes. This is indicated by the phase traces jumping above and below the 90-degree line and the presence of probe and sample damage. **f,** Trace and retrace lines in repulsive mode. When the setpoint is small, the probes are in the fully repulsive mode. In this mode, there is a high risk of probe and sample damage.

**Define the parameter space and reward function**

Below we discuss the relationship between different SPM modes, scan quality, safety of the sample and probe, and the tip-sample distance. In SPM, the tip-sample distance is controlled by the ratio between the setpoint amplitude ($A_{setpoint}$) and free-air amplitude ($A_{drive}$, also known as free amplitude) so the discussion of the physical mechanisms will be guided by the ratio of setpoint amplitude and free-air amplitude.

When the tip is far away from the sample surface (large $A_{setpoint}/A_{drive}$), there are no forces between tip and sample. Consequently, the probe is oscillating far above the sample surface instead of following the surface profile, leading to phases close to the free-air phase and random height traces as shown in Figure 2c, subjected to either small drift or other noise in the signal.

As the tip gets closer to the sample surface (smaller $A_{setpoint}/A_{drive}$), SPM works in the attractive mode and there are typically long-range attractive forces arising between the tip and surface. This attractive force will decrease the amplitude of the cantilever to the setpoint amplitude at a fixed tip-sample distance, as shown in Figure 2d. As a result, the trace and retrace lines follow closely the sample surface profile with reasonably high spatial resolution and the phases are above the free-air phase. In addition, the force between the tip and sample is kept at a small, net-attractive value and that helps minimize tip and sample damage. This mode is the goal for this work.

When the tip gets even closer to the sample surface (smaller $A_{setpoint}/A_{drive}$), SPM jumps between attractive mode and repulsive mode as the interaction between the tip and sample jumps between the long-range attractive forces and short-range repulsive forces. Consequently, there are sudden jumps in the phase traces across the free-air phase in this crossover mode, as shown in Figure 2b. A previous study showed that the transition between attractive and repulsive mode can result in very high transitory forces between the tip and the sample [33] .As a result, the trace and retrace lines match poorly due to the high transitory forces when the phase jumps occur in Figure 2e.

When $A_{setpoint}/A_{drive}$ is further decreased, the interaction between the tip and sample is more likely to be completely dominated by the short-range repulsive forces. In this mode, the phases are below the free-air phase most of time. The spatial resolution is suppressed and the tip wears off quickly due to the large forces between the tip and sample, as shown in Figure 2f.

Based on these physical considerations, we suggest the reward function based on four components derived from the trace and retrace scan lines of the height and phase channels. The height alignment term quantifies the agreement between the trace and retrace lines in the height channel. The phase term measures if the probe is operating in the attractive mode by detecting the presence of phase angles below the free-air phase. The distance term favors lower probe heights to make sure the probe is as close to the sample surface as possible. Finally, the contrast term tries to maximize the amount of information acquired by favoring a large contrast in the height channels. Assembling all these components together, we define the reward function based on a pair of trace and retrace lines of height and phase channels as following equation:

$$\text{Rewards} = -\log \left( \sum \frac{|h_{trace} - h_{retrace}|}{h_{trace} + h_{retrace} - \min(h_{trace}, h_{retrace})} \right) \times \log(N_{\theta < \theta_{free}} / N_{\theta > \theta_{free}}) \times \log \left( \frac{h_{min} - h_{global\,min}}{h_{max} - h_{min}} \right) + \log \left( \frac{\sigma_{traces}}{h_{max} - h_{min}} \right) \quad (1)$$

where $h_{trace}$ and $h_{retrace}$ are the trace and retrace height scan lines with linear fitted background removed, $N_{\theta<\theta_{free}}$ and $N_{\theta>\theta_{free}}$ are the total number of pixels in the phase scan lines smaller (repulsive mode) and greater (attractive mode) than the free phase angles, $h_{max}$ and $h_{min}$ are the highest and lowest probe position of current trace and retrace height lines, respectively, $h_{global\ min}$ is the global lowest probe position computed based on all the acquired scan lines, $\sigma_{traces}$ is the standard deviation of the height lines to represent the contrast. For all the terms inside log operator, the values are capped minimally at logarithmic constant e to prevent the log terms generating negative values.

By normalizing the terms inside log operators with either the largest height difference or the height traces at each pixel, the effect of height variation across the whole scan area is minimized. To remove the latency between the neighboring trace and retrace that are usually induced by phase lag inside the instrument controller, we performed a shift match before computing the rewards. We shifted the retrace with respect to the retrace, and found the optimal shift value by minimizing the absolute difference between the two trace. Please see the attached example Jupyter notebook for details.

When the probe is far away from the sample surface and only very weakly interacting with the sample, the traces will be dominated by cantilever sensor noise. Sensor noise will generally have little or no correlation between trace and retrace. The resulting height traces agree poorly with each other and the distance between the probe and sample is large as shown in Figure 2c. Therefore, both the height alignment term and distance term give bad reward values to signal a bad scan. In the attractive mode, every component in our reward function gives good rewards as the height traces align well, the phases are consistently above the free-air phase, the probe is close to the sample surface, and there is a reasonable amount of contrast, as shown in Figure 2d. In the crossover mode as shown in Figure 2e, the phases below the free-air phase leads to a bad reward in the phase term. In the repulsive mode as shown in Figure 2f, most of the pixels in the scan lines show phases below the free-air phase, giving a very small reward to prevent the SPM from working in this mode.

Overall, the reward function is designed to distinguish between a set of good topography scanning parameters in the attractive mode from bad parameters in other modes. In addition, it also leaves the flexibility to adjust the weight of different components to fine-tune for different applications.

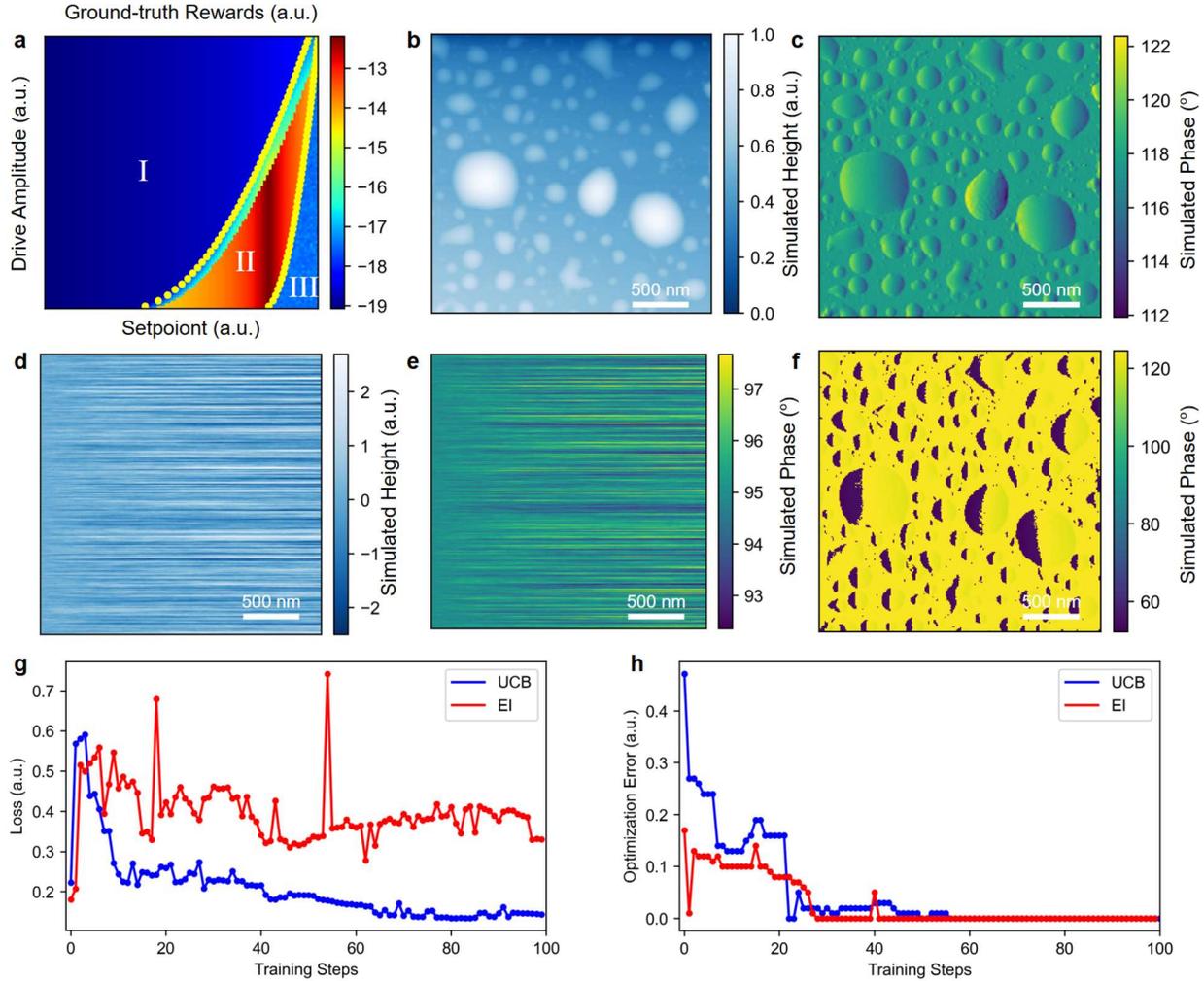

**Figure 3. Benchmark reward functions with SPM scan simulator. a,** ground-truth reward function simulated on a dense grid of drive amplitudes and setpoints. Two dashed yellow curves divide the map into three parts. Region I: the setpoint / drive is too small and probe is working in the repulsive mode; region II: the probe is operating in the attractive mode as the setpoint / drive is large enough; region III: the setpoint / drive is so large that the scan is dominated by random noise. **b,** ground-truth height map used in this simulation. This map is acquired on a water droplet sample experimentally. **c,** simulated phase map based on b with an ideal set of scanning parameters in the region II. **d-e,** simulated (d) height and (e) phase maps with scanning parameters in the region III. The setpoint / drive is so large that the scan is completely dominated by noise. **f,** simulated phase map with scanning condition in between region I and II. The transition between attractive and repulsive modes induces large phase jumps in the areas with large slopes in the height map. **g,** comparison of learning loss between acquisition functions of upper confidence bound (UCB) and expected improvement (EI). At each step, the loss is computed as the absolute difference between the rewards predicted by the learned BO model and the ground-truth rewards measured on a dense grid as shown in a. The UCB and EI simulations both start with the same set of five random initial seeding points. **h,** optimization error is computed as the Euclidean distance between the predicted optimal scanning parameters at each training step and the true optimal parameters given by the grid-search in a.

**Benchmark the performance of reward function with a simulator**

In the real active learning autonomous experiments, we usually don't have access to the full picture or the ground truth, which imposes difficulty in benchmarking the performance of autonomous experiments. For example, in the automated optimization of SPM in the tapping mode, we only have access to the rewards computed from all the acquired scanning traces, which are limited to a small fraction of the entire parameter space. At the same time, we do not have access to the "true" surface topography. Hence, defining the benchmarking approaches similarly to how it is done in the classical ML community on human labeled data sets is impossible.

To resolve this problem, we designed an open-source SPM scanning simulator, SpmSimu, to benchmark the performance of our reward-based automated optimization workflow. It simulates the trace and retrace scanning maps of height and phase channels based on ground-truth height patterns, tip shape kernels, and scanning parameters provided by users. The ground-truth patterns can either be artificially generated height profiles or real topographic profiles acquired experimentally. This simulator is designed to capture the realistic effect of different tip conditions including tip radius and multiple tips, scanning parameters including drive amplitude, setpoint, proportional (P) gain, integral (I) gain, PI buffer size, z-piezoelectric speed, and scanning speed (xy-piezoelectric speed). In addition, it's capable of generating realistic phase response according to the simulated real time tip-sample distance and the applied drive amplitude.

To verify that the simulator captures the characteristics of real SPM scanning in the tapping mode, simulations with different drive amplitudes and setpoints were performed. As shown in Figure 3a, the parameter space is divided into three regions. In region I, the $A_{setpoint}/A_{drive}$ is too small and the probe is working in the repulsive mode. As a result, the reward is very small signaling a bad scan quality. In region II, the $A_{setpoint}/A_{drive}$ is ideal and the probe is operating in the fully attractive mode. The simulated phase map in Figure 3c resembles the behavior of probes in a real experiment as shown in Figure 2d. In region III, the $A_{setpoint}/A_{drive}$ is so large that even the random instrument noise can trigger the setpoint, which leads to totally random scanning traces shown in Figure 3d-e. This behavior corresponds to the "not on surface" case in the real experiment shown in Figure 2. Besides, crossover between attractive and repulsive modes in Figure 2e is reproduced by our simulator running with scanning parameters in-between region I and II, as shown in Figure 3f. The implementation of the SPM scanning simulator and its simulation of double-tip and other commonly observed experimental artifacts can be found in the Appendix and Figure S1.

To benchmark our proposed reward function, a dense full-grid search in the parameter space was first performed. For easy visualization of the results, we have selected the drive amplitude and setpoint as our parameters to optimize. The simulation was simulated based on the water droplet topography profile acquired experimentally with an ideal tip as shown in Figure

S1-a, and all the other parameters were set to their empirically chosen optimal values. In the grid-search, trace and retrace scan maps of height and phase channels were simulated with the given drive amplitude and setpoint, from which the reward was computed based on the full-maps of height and phase using equation (1). The ending result is a map of rewards in the 2D parameter space, as shown in Figure 3a.

Two simulations of automated optimization of scanning parameter experiments were performed based on the simulator. In the simulation with the acquisition function of UCB, 5 random initial seeding points were selected from the 2D parameter space, and their corresponding reward values were computed based on the height and phase maps simulated with these selected seeding parameters. The training process consisted of iterative steps, which starts with fitting a Gaussian Process (GP) model based on all acquired reward values. Then the UCB acquisition function is computed based on the prediction and uncertainty of the GP model, from which the next pair of parameters to try is determined. The EI simulation starts with the same 5 initial seeding parameters, but it uses the acquisition function of EI to determine the next parameters to try. Both simulations consist of 5 initial seeding points and 100 training steps. In the UCB simulation, the acquisition function parameters are chosen to favor the tendency of exploration over exploitation.

The loss function at each step is computed by summing the absolute difference between the reward distribution predicted by the GP model and the ground truth from the grid-search of the rewards, which measures the accuracy of the GP model in describing the full picture of reward distribution over the entire parameter space. As shown in Figure 3g, the acquisition of UCB has a better performance in terms of getting a more accurate full-picture description of the reward distribution, which agrees with the chosen exploration policy for the UCB simulation.

To compute the optimization error, we first find the ground-truth best scanning parameters by finding the scanning parameters corresponding to the maximum reward in the grid-search shown in Figure 3a. At each training step, the optimization error is computed as the Euclidean distance between the optimal parameters predicted by the GP model and the ground-truth best parameters. Therefore, the optimization error describes how close each model is able to predict the best scanning parameters. As shown in Figure 3h, the acquisition of EI has a better performance than the UCB. It has a closer initial guess, a faster decay, and smaller ending steps. The videos of the training process can be found in the attached media files.

This SPM scanning simulator can accurately capture the different scanning modes observed in the real experiments and describe the effects of different scanning conditions. We note that while SPM model here is basic, it is unknown for the optimization algorithm and therefore we expect that if trained on unknown dynamic model, similar algorithm will work for realistic environments. It provides a cheap but effective way to simulate realistic scan images at different tip and scanning conditions, which can be used for user training, development of instrument automation and microscopy-related machine learning applications. The benchmarking results show that our proposed reward function can capture the characteristics of good scans over bad scans, and the BO automated optimization workflow can quickly discover the optimal scanning parameters.

**Reward based optimization and results**

We employed Bayesian optimization (BO) to optimize the scanning parameters, focusing on two key parameters: free-air amplitude and setpoint, under consistent conditions. In addition to the physical knowledge about SPM that we encoded into the reward function, we found it necessary to guide the optimization process with human empirical knowledge, especially on defining the parameter space.

To set up a high-resolution scan in the attractive mode, human operators usually start with a small driving voltage and a large setpoint that ensures gentle force between the probe and sample. The smallest free amplitude can be estimated by the roughness of the sample in the given scan size. Then the setpoint is decreased from a large starting value until the trace and retrace lines agree with each other and thus follow the corrugation of the sample surface. Once the free-air amplitude and setpoint are determined, human operators can also increase the integral gain (I gain) as high as possible without the PI loop becoming unstable and oscillatory.

Initially, we defined the parameter space based on the type of probe and the estimated roughness of the sample in the given scan size. For instance, with the Tap300G probe and $TiO_2$ nanoparticles, the free-air amplitude was constrained between 0.15 nm and 5.9 nm, and the setpoint ranged between 10% and 90% of the free-air amplitude. The optimization process and results for $TiO_2$ nanoparticles are shown in Figure S2.

After verifying the functionality of our reward-driven optimization workflow on the $TiO_2$ nanoparticles, we switched to a water droplet sample to challenge our reward-driven workflow. The water droplet on a mica substrate serves as an excellent testing sample for our autonomous optimization workflow. First, the image quality of water droplets is highly sensitive to the tip-sample distance. If the tip-sample distance is too large, the probe cannot accurately follow the height profile of droplets. Conversely, if the tip-sample distance is too small, a liquid bridge forms between the tip and the droplet, resulting in distortions in the image. More importantly, tuning the scanning parameters for water droplets is challenging even for experienced SPM operators, as the optimal setpoint amplitude and free-air amplitude are confined to a narrow region in the parameter space.

Once the parameter space was established, we randomly sampled 10 points within this space to serve as the initial seeding points. The microscope was then controlled via code to measure the scan lines at these parameters, from which the reward function could be computed, as illustrated in Figure 4a. Using the reward values from these seeding points, we fitted an initial Gaussian process (GP) model. This model predicted the distribution and uncertainty of the reward function across the entire parameter space. Subsequently, the acquisition of expected improvement (EI) was computed based on these predictions and uncertainties as shown in Figure 4b-c, determining the next set of parameters to be measured by maximizing the EI acquisition function. To account for human empirical knowledge of starting the optimization with small free-air amplitude and large setpoint, we masked the acquisition function in the parameter space to enhance the preference on the parameters with smaller free-air amplitude and larger setpoint, as shown in Figure 4c. The scanning parameters were then updated to the new set without interrupting the scan. The reward value for these new parameters was calculated from the newly acquired trace and retrace scan lines. The new parameters-reward pair was added to the previously collected data to retrain and update the GP model. This iterating process of measuring and fitting was repeated until the maximum number of exploration steps was reached or the scan quality became satisfactory.

Despite these challenges, we were able to find the optimal scanning parameters that give high-quality TM scans of water droplets with our autonomous workflow as shown in Figure 4e-f The sharp images of small droplets indicate a high spatial resolution in the image. The smooth image of the large droplets in the height map together with the phase map above the free-phase reveals that the probe works in the net-attractive mode across the whole image and closely follows the height profile of the droplets without distortion.

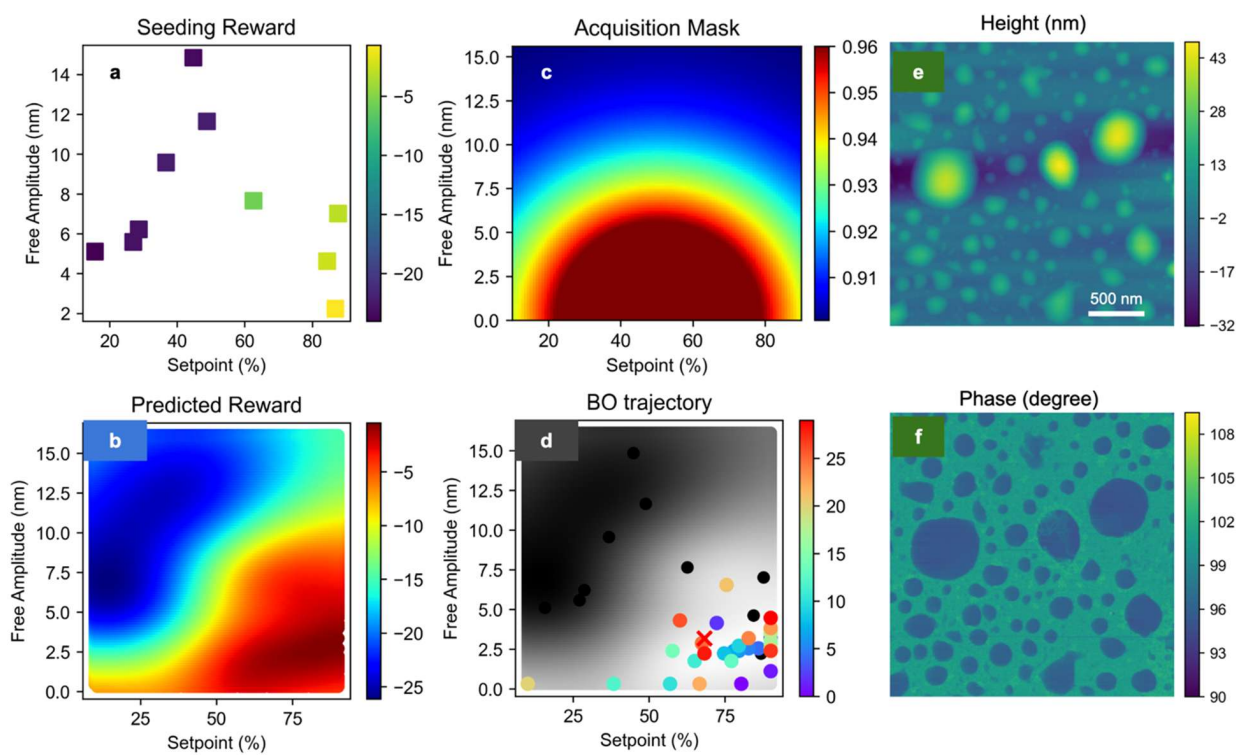

**Figure 4. Optimization for more challenging materials: water droplet. a,** Initial seeding of 10 points and their reward values. **b,** predicted distribution of reward function in the full parameter space after 30 BO exploration steps. **c,** Acquisition mask for guiding the optimization process with human empirical knowledge. By multiplying the EI acquisition function with this acquisition mask, the parameters with lower free-air amplitude and larger setpoint are preferred for the next iteration. **d,** Bayesian optimization trajectory of the seeding and exploration (see Figure S1 for the detailed optimization data associated with the BO trajectory). The optimized scanning parameters are marked by a red cross. **e,** Topography map of water droplet (with $CaCl_2$ salt saturated solvents) taken with the optimized scanning parameters in **d**. **f,** Phase map taken together with **e** shows the full image is taken in the attractive mode.

**Conclusion**

Our reward-driven automatic optimization of SPM in tapping mode represents a significant advancement for the SPM community. This method not only drastically reduces the instrument time required to tune different probes on various samples but also ensures the acquisition of high-quality and highly reproducible SPM images. This automation increases the application and accessibility of SPM by eliminating the need for extensive manual tuning. Our study lays the groundwork for automating more complex SPM tasks, such as automating more advanced imaging modes and achieving atomic resolution in ambient environments.

Moreover, the broader applicability of this reward-driven optimization workflow across the microscopy community has the potential to transform traditionally complex human-led tasks into streamlined, machine learning-based optimization processes. This shift is crucial for the further automation of microscopy, addressing one of the last significant barriers to fully automated microscopes.

Finally, we anticipate that our workflow will have a profound impact on open science in microscopy. While sharing microscope data alone has provided limited benefits due to the diversity of tasks and applications, the sharing of reward functions introduces a new paradigm. This approach fosters a more integrated and collaborative community, enhancing the reproducibility and efficiency of microscopy research.

This comprehensive solution not only democratizes access to high-quality SPM imaging but also propels the field towards a future where automated, intelligent systems are the norm, significantly advancing both research and industrial applications.

**Methods**

**Technical settings**

For the implementation in this paper, we made it light enough to run on a local computer with central process unit (CPU) only. Optimization of scanning parameters in the tapping mode requires an interface library to directly control the SPM and availability of sufficiently high computational power to support the optimization algorithms. Previously, we have developed a Python interface library to control the automation of a Jupiter SPM system manufactured by Oxford Instrument Asylum Research from a local computer without GPU. Connection with HPC will allow more complicated optimization scheme and we have solutions for that [34]. This library not only enables operating the SPM system remotely with code the same way as human operators, but it also has access to the intermediate data like trace and retrace scan lines in all the channels.

**Samples**

The (CrVTaW)$_x$Mo$_{1-x}$ thin film was grown via dc magnetron co-sputtering from a 50 mm diameter Mo and an equiatomic CrVTaW target at 500°C substrate temperature. The system was pumped to ~ 3x10$^{-7}$ Torr and backfilled with Ar to 5 mTorr and the sputtering powers (200 W for CrVTaW and 100 W for Mo) were adjusted to give approximately equivalent sputtering rates (10 nm/min determined via x-ray reflectance) of the two targets at the substrate center. The pseudo binary (CrVTaW)$_x$Mo$_{1-x}$ composition varies from 15 <x <88 at. % across the 100 mm diameter substrate with a roughly linear composition gradient.

The PbTiO$_3$ (PTO) thin films were grown on La$_{0.7}$Sr$_{0.3}$MnO$_3$ (LSMO) buffered (110)-oriented SrTiO$_3$ (STO) single crystal substrates using pulsed laser deposition (PLD) with a KrF excimer laser ($\lambda$ = 248 nm). The LSMO/PTO layers were deposited at temperatures of 700 °C/690 °C with oxygen pressures of 100 mtorr/150 mtorr, respectively. After deposition, the samples were cooled to room temperature under an oxygen pressure of 700 Torr. The thicknesses of the PTO and LSMO layers are approximately 150 nm and 30 nm, respectively.

Pb$_{0.995}$(Zr$_{0.45}$Ti$_{0.55}$)$_{0.99}$Nb$_{0.01}$O$_3$ films were grown by pulsed laser deposition using a KrF excimer laser from a ceramic target onto a SrRuO$_3$-electroded (001) SrTiO$_3$ single crystal. The SrRuO$_3$ film was grown from a target from Kojundo Chemical Lab. Co. Ltd., using a laser energy density of 1.5 J/cm$^2$, a substrate temperature of 660°C, an oxygen pressure of 120 mTorr, a target-to-substrate distance of 6.7 mm, and a frequency of 5 Hz. The SrRuO$_3$ film thickness was around 50 nm. The PZT film was grown from a target with 20% excess PbO to compensate for lead loss during growth, using a laser energy density of 1.5 J/cm$^2$, a substrate temperature of 630°C, an oxygen pressure of 120 mTorr, a target-to-substrate distance of 6.2 mm, and a frequency of 5 Hz. The PZT film thickness was around 147 nm.

TiO$_2$ NPs were synthesized via a modified solvothermal method. Titanium tetraisopropoxide (TTIP, 99.95%, Sigma Aldrich) served as the titanium precursor, and anhydrous ethanol was utilized as the solvent. In a standard synthesis, 0.1 M of TTIP was combined with 50 mL of ethanol and stirred uniformly for one hour. Subsequently, a mixture of 50 mL ethanol and deionized water in a 1:1 ratio was added to the solution, followed by vigorous stirring for another hour. To induce rapid hydrolysis, a few drops of pH 4.0 H$_2$SO$_4$ were introduced. The resulting white solution was then placed in a tightly sealed reagent bottle and subjected to solvothermal growth at 90°C for 12 hours. The final products were centrifuged and washed several times with de-ionized water to eliminate any bound sulfate ions. The resulting NPs were dried and annealed at varying temperatures (400 °C) for one hour. These obtained NPs were characterized by X-ray diffraction and scanning electron microscopy for their structure evaluation.

The Cs$_3$Bi$_2$I$_9$ (CBI) microcrystals (MCs) were synthesized through a solvent-free mechanochemical activation method, as detailed elsewhere [35]. In this procedure, the precursors cesium iodide (CsI; Aldrich, 99.999% ultra-dry, metals basis), bismuth iodide (BiI$_3$; Sigma-Aldrich, 98%), were ground for 15 minutes under ambient conditions using a mortar-pestle. The resultant powder was then washed three times with a mixture of dimethyl formamide (DMF) and dimethyl sulfoxide (DMSO) solution. Afterwards chloroform (CHCl$_3$) was utilized as an antisolvent to obtain orange microcrystals. The MCs were characterized by X-ray diffraction and scanning electron microscopy for their structure evaluation.

The preparation of CaCl$_2$-solution droplets on mica substrate follows the recipe described in [22].

**SPM setup**

All the tapping mode SPM measurements were performed on a Jupiter SPM system manufactured by Oxford Instruments Asylum Research. The probes tested include ASYELEC.01-R2 and AC 160TS-R3 from Asylum Research, Multi75G and TAP300G from BudgetSensor, and AD-2.8-AS diamond probe from Adama Innovations.

**Code Implementation**

Botorch [36] was utilized as the Bayesian optimization framework, employing the Expected Improvement (EI) acquisition function. The surrogate Gaussian Process (GP) model was incorporated using gpytorch [37]. In the optimization of two scanning parameters described in this paper, we sampled the reward function in a parameter space with 100 x 100 pixels. All the parameters are normalized to be within (0, 1).

AESPM is an open-source SPM-Python interface library. It can be found in the following link with detailed examples and tutorial notebooks: https://github.com/RichardLiuCoding/aespm

SpmSimu is an open-source SPM scanning simulator. It can be found in the following link with detailed examples and tutorial notebooks: https://github.com/RichardLiuCoding/spmsimu

To make sure readers can reproduce the results shown in this work, we have prepared an open-source Jupyter notebook in this link: https://github.com/RichardLiuCoding/spmsimu/blob/main/spmsimu/notebooks/TM_BO_Benchmarking_SpmSimu_v1.ipynb. This notebook can be run either online on Colab or locally after downloading. It contains all the codes to reproduce the benchmarking results in Figure 3 and Figure S1. An example of a complete workflow of automated optimization of drive amplitude and setpoint based on the SpmSimu simulator is also provided. Another example shows the optimization of three parameters.

With AESPM, SpmSimu and example automated optimization workflow notebook provided, readers should be able to reproduce our results based on the simulator, or modify the provided codes to make the workflow work on their own instruments.

**Acknowledgement**

The development of automated workflow for SPM optimization (YL, SVK) was supported by the Center for Advanced Materials and Manufacturing (CAMM), the NSF MRSEC center. The combinatorial library growth (RE, PDR) was supported by the National Science Foundation Materials Research Science and Engineering Center program through the UT Knoxville Center for Advanced Materials and Manufacturing (DMR-2309083). $TiO_2$ nanoparticles and CBI and MBI microparticles synthesis (A. D.) are supported by individual postdoctoral funding program German Academic Exchange Service (Deutscher Akademischer Austauschdienst: DAAD-PRIME2023). PTO (110) growth (JCY, YCL) acknowledged the financial support from National Science and Technology Council (NSTC) in Taiwan, under grant no. NSTC 112-2112-M-006-020-MY3. M.A. acknowledges support from the National Science Foundation, Award Number 2043205.

## Supplementary Materials

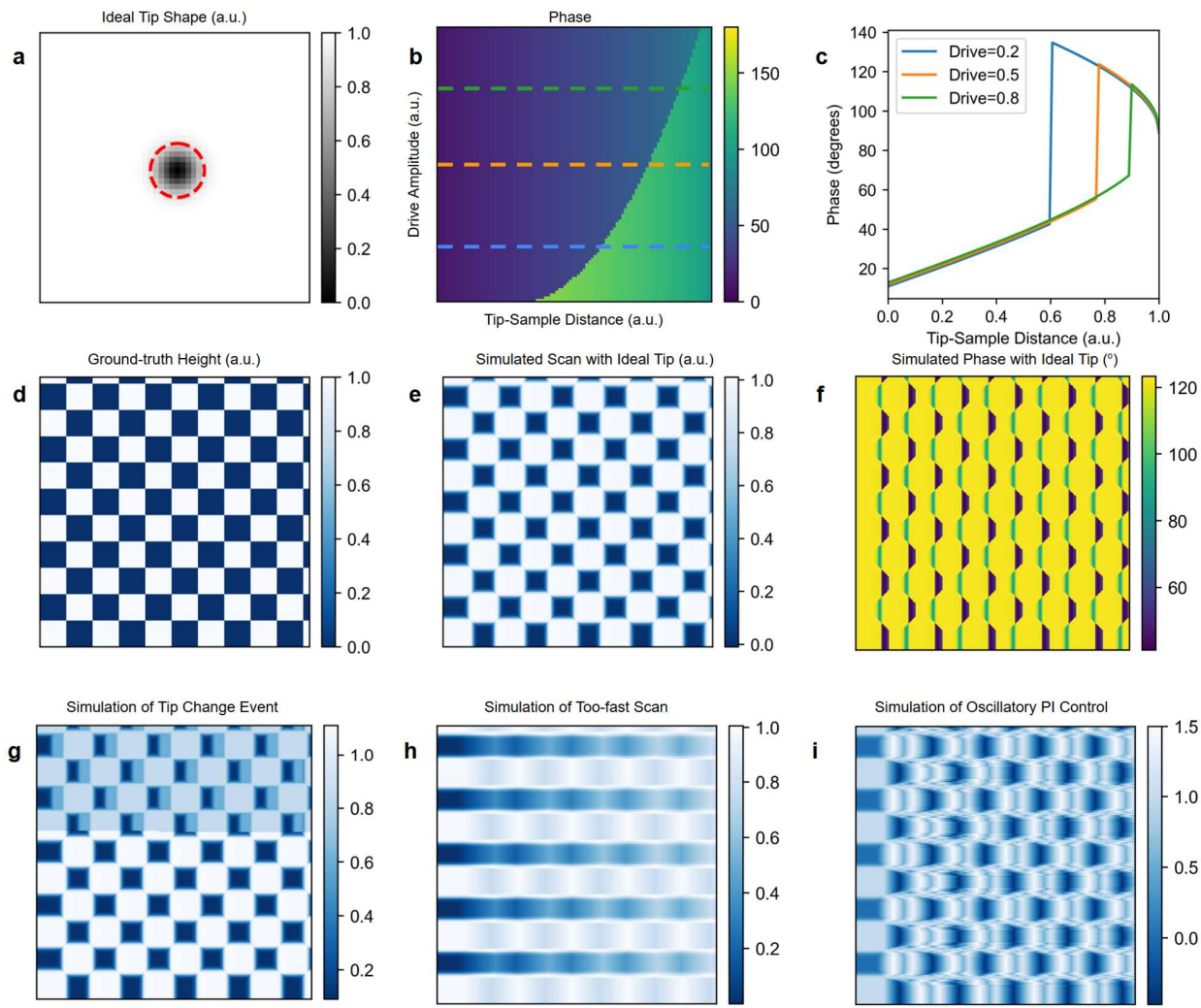

**Figure S1. SPM scanning simulator. an example** of an ideal tip kernel with a Gaussian profile. The red dashed circle represents the half-max width of the probe and controls the resolution of the probe. **b-c,** phase-distance response at different drive amplitudes. Three example response curves at selected drive amplitudes are plotted in c. **d,** ground-truth checkerboard pattern. **e,** simulated scan height map based on pattern in d, with ideal tip profile in a, and with optimal scanning parameters. **f,** corresponding phase map shows the change of phase at the step edges due to sudden change of sample height. **g,** simulation of a tip change event. After the tip change, the tip becomes double and shorter compared to before the tip change. **h,** simulated effect of scanning with too-fast speed. The tip is moving so fast in the lateral direction that the z-piezoelectric driver is not able to respond to changes in the sample height in time. **i,** simulated effect of oscillatory PI controlling loop. When the I gain is too large, the PI controlling loop itself becomes unstable and there will be ripples showing up in the scans.

## I. Implementation of scanning probe microscope scanning simulator

The SpmSimu simulator consists of two parts: height simulator and phase simulator. In the height simulator, the tip is modeled as a kernel in a smaller square box compared to the entire map as shown in Figure S1a. Different tip conditions can be modelled by controlling the width of peaks, number of peaks and their separations. When the kernel is scanning across the map, the closest distance between the kernel and map in the overlapping area is used as the distance push down at the center of the kernel, which can be used to calculate the measured sample height at the center of the kernel. This method guarantees that the dilation effect is precisely captured no matter how the tip and sample conditions change [38].

The effect of P gain, I gain, scan speed, and z-speed are realized through the simulated PI control loop. Here, the next tip height is determined with the following equation:

$$z_{tip}^{(i)} = z_{tip}^{(i-1)} + P \times (s - s_0) + I \times \sum_0^t (s - s_0) \delta t \qquad (2)$$

where $z_{tip}^{(i)}$ is the expected tip height at position $i$, $z_{tip}^{(i-1)}$ is the measured tip height at position $i - 1$, $s$ is the measured optical sensor amplitude (LVDT), P is the integral gain constant, I is the integral gain constant, $\delta t$ is the time between two neighboring pixels which is set to 1 for simplicity in the simulator, $t$ represents how many previous data will be accounted for in the integral term and $s_0$ is the setpoint amplitude. If we assume that the setpoint amplitude $s_0$ corresponds to a constant tip-sample distance $z_0$, the equation (2) can be rewritten as:

$$z_{tip}^{(i)} = z_{tip}^{(i-1)} + P \times \left(z_{tip}^{(i-1)} - z_{sample}^{(i-1)}\right) + I \times \sum_0^t \left(z_{tip}^{(i-1)} - z_{sample}^{(i-1)}\right) \delta t \qquad (3)$$

where $z_{sample}^{(i-1)}$ is the height of the ground-truth pattern at position $i$ [4, 26].

However, due to the finite z-piezoelectric driver speed, the actual tip height at the next pixel $i$ can be different from the expected value given by the PI algorithm, especially when the scan speed is fast and/or the sample height is changing quickly. Therefore, the change in the tip height between neighboring pixels will be capped at the maximum allowed height change, which is computed based on the scan speed and z-speed. In the end, the measured sample height is computed based on the measured tip height instead of the expected tip height. This dependence on the scan speed and z-piezoelectric speed is described by the following equation:

$$\begin{aligned} z_{tip}^{(i)} &= z_{tip}^{(i)} \text{ if } |z_{tip}^{(i)} - z_{tip}^{(i-1)}| < \Delta z_{max} \\ &= z_{tip}^{(i-1)} + sign\left(z_{tip}^{(i)} - z_{tip}^{(i-1)}\right) \times \Delta z_{max} \text{ if } |z_{tip}^{(i)} - z_{tip}^{(i-1)}| \geq \Delta z_{max} \end{aligned} \quad (4)$$

here $\Delta z_{max} = v_{z-pie} \times \frac{\Delta d}{v_{scan}}$ represents the maximum z-piezoelectric movement distance between two neighboring pixels, $v_{z-pie}$ is the z-piezoelectric movement speed, $v_{scan}$ is the xy-scan speed and $\Delta d$ is the distance between two neighboring pixels.

The phase map is generated based on the difference between the measured sample height and the ground-truth height map. Once the drive amplitude is set, the phase – distance relation will be a horizontal cut in the Figure S1b, based on which the phase map can be calculated. Here to simulate the realistic phase response, we have assumed that the attractive – repulsive crossover distance is a function of setpoint as shown in Figure S1b-c, and the threshold values of A$_{setpoint}$/A$_{drive}$ for the tip to be "not on surface" is also a function of setpoint, as shown in Figure 3a.

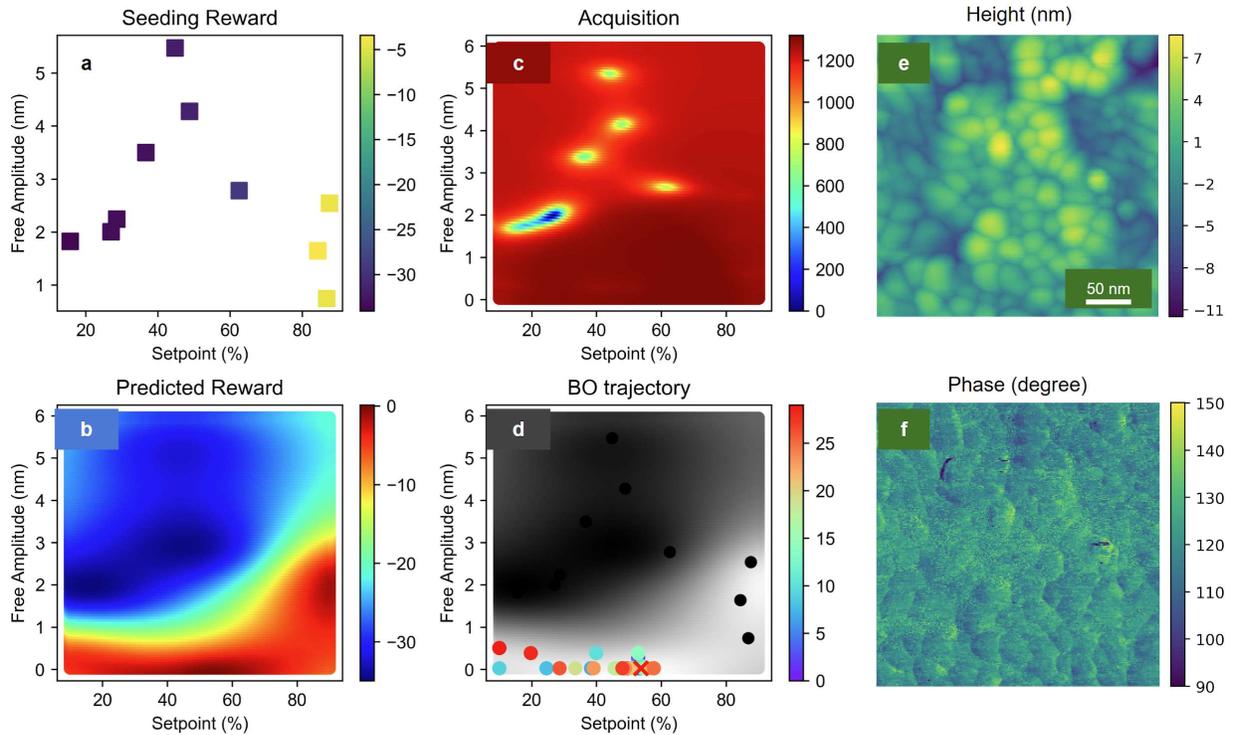

**Figure S2. Autonomous tuning of SPM with BO on TiO$_2$ nanoparticles. a,** Initial seeding of 10 points and their reward values. **b**, Predicted distribution of reward function in the full parameter space after 30 BO exploration steps. **c,** Acquisition function of log expected improvement (EI). This is computed based on the prediction and uncertainty of the reward function shown in **b** and is used to determine which combination of parameters to measure for the next iteration. **d,** Bayesian optimization trajectory of the seeding and exploration. The optimized scanning parameters are marked by a red cross. **e,** Topography map of TiO$_2$ nanoparticles taken with the optimized scanning parameters in **d**. **f**, Phase map taken together with **e** shows the full image is taken in the attractive mode.

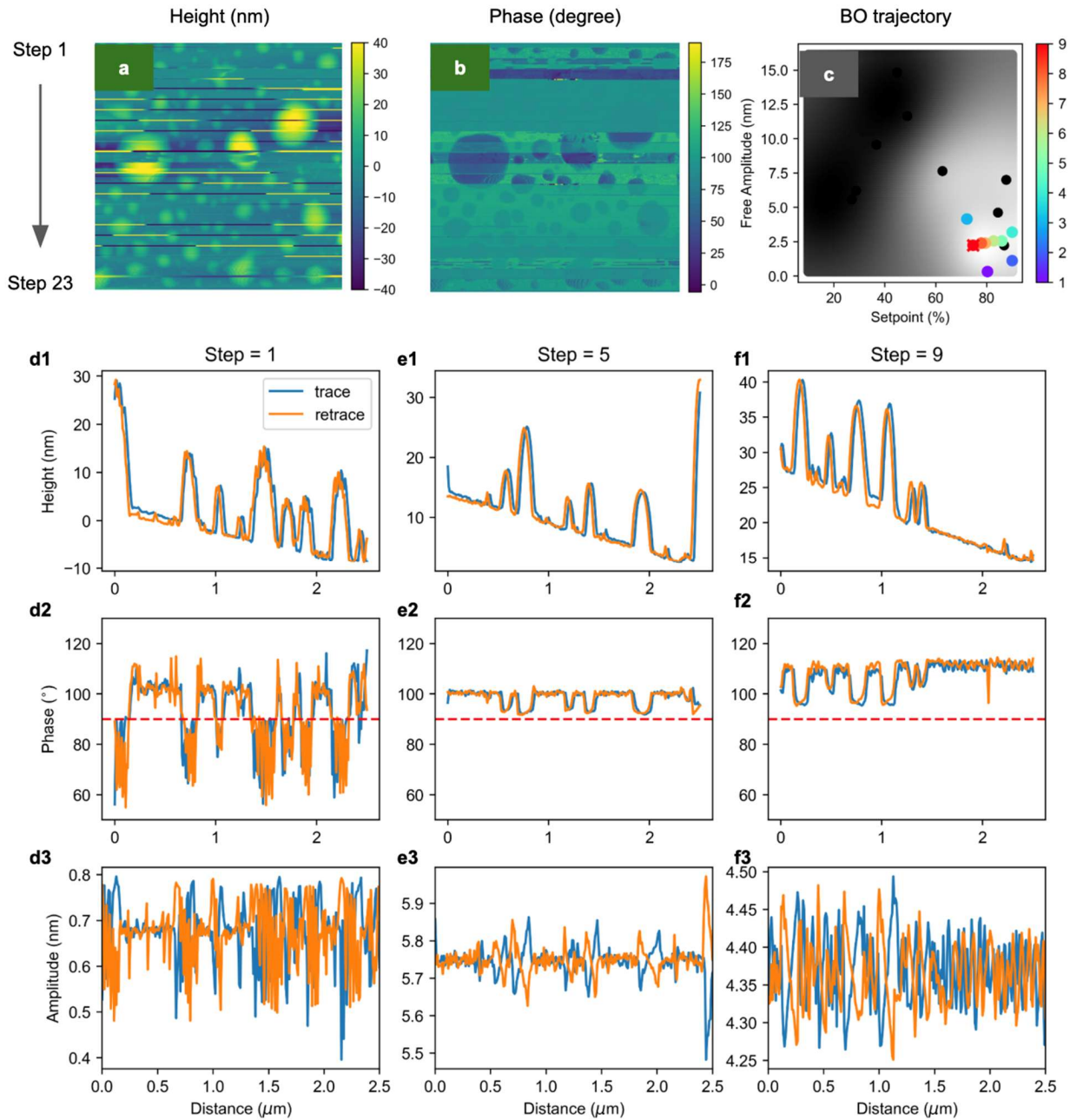

**Figure S3. Visualization of the training process on water droplets. a,** Topography map acquired on the droplet sample during the training process. In this map, each "jumping line" represents an update of scanning parameters and there are 23 steps in this single scan. After the scanning becomes stable, five subsequent trace and retrace scanning lines of height, phase and amplitude channels will be recorded to compute the reward function for currently used parameters. **b,** corresponding phase map shows the evolution of phase angles during the training process. **c,** BO trajectory shows how the workflow explores the parameter space. **d1-d3,** the trace and retrace scanning lines of **d1** height, **d2** phase, and **d3** amplitude channels at step 1. Similar scanning lines for **e1-e3** step 5 and **f1-f3** step 9 are also plotted. The images taken with the optimized parameters are shown in the Figure 4.

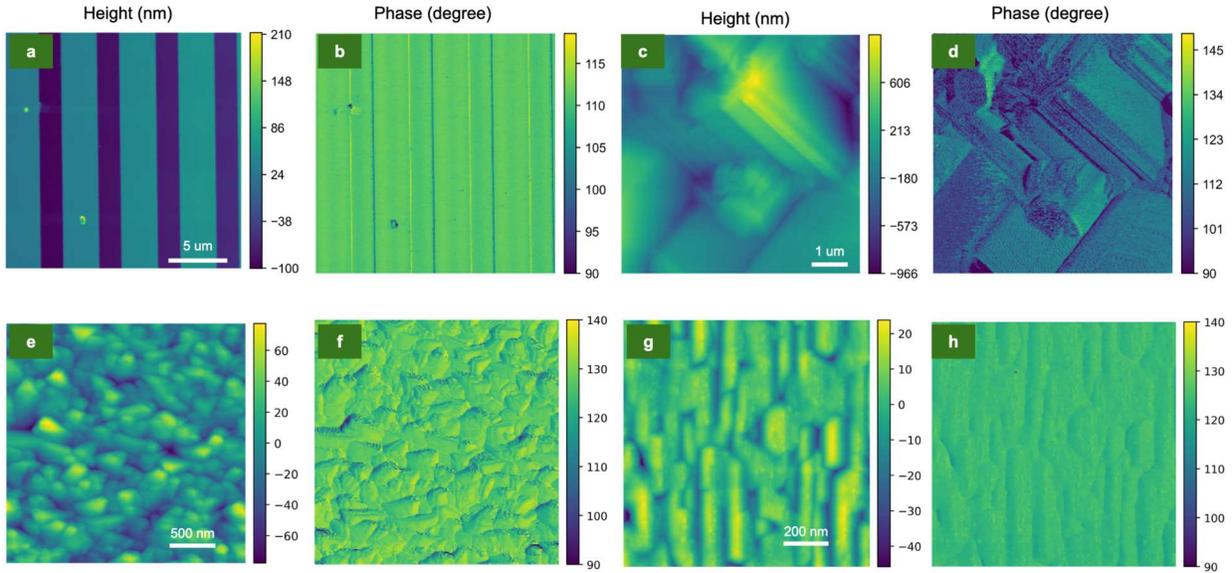

**Figure S4. Testing the workflow on different samples. a,** Topography map acquired with the optimized scanning parameters given by our automated workflow on height calibration grating sample from Asylum Research. **b,** corresponding phase map taken on the calibration grating sample. Similar maps acquired on **c-d** CBI nanoparticles **e-f** $(CrVTaW)_xMo_{1-x}$ alloy **g-h** $PbTiO_3$ ferroelectric thin-film.

## II. Detailed analysis of BO exploration trajectories

Surprisingly, the BO exploration trajectory in Figure S3 shows that our autonomous workflow emulates how experienced human operators optimize a microscope: after a few random exploratory steps, BO explores the scanning parameters along a pathway from small free-air amplitude and a large setpoint amplitude at step 3 to decreased free-air amplitude and setpoint amplitude at step 9.

## III. Universality of the optimization

To verify the universality of our reward-driven autonomous optimization workflow, we tested the same workflow over different probes and samples. Samples tested include height calibration grating sample from Asylum Research, $PbTiO_3$, $(CrVTaW)_xMo_{1-x}$, CBI microparticles, $TiO_2$ nanoparticles, water droplets with $CaCl_2$ salt.

Topography maps in Figure S4 were taken with scanning parameters optimized by our autonomous workflow and showed that our workflow can give optimal parameters for a wide range of samples with different scanning sizes, sample roughness and sample hardness.